\documentclass[a4paper,conference]{IEEEtran/IEEEtran}
\def\ps@headings{
\def\@oddhead{\mbox{}\scriptsize\rightmark \hfil \thepage}
\def\@evenhead{\scriptsize\thepage \hfil \leftmark\mbox{}}
\def\@oddfoot{}
\def\@evenfoot{}}
\makeatother

\usepackage{pgfplots}
\usepackage{color}
\usepackage{subfig}
\usepackage{graphicx}
\usepackage{cite}
\usepackage{amssymb,amsfonts}
\usepackage{url}
\usepackage[utf8]{inputenc}
\usepackage{tabularx}
\usepackage{hhline}
\usepackage{multirow}
\usepackage{wrapfig}

\usepackage[outline]{contour}

\usepackage{pifont}
\usepackage{tikz}

  \usetikzlibrary{arrows,shapes,calc,positioning,shadows,trees,patterns,decorations}
  \tikzset{
    ncbar angle/.initial=90,
    ncbar/.style={
        to path=(\tikztostart)
        -- ($(\tikztostart)!#1!\pgfkeysvalueof{/tikz/ncbar angle}:(\tikztotarget)$)
        -- ($(\tikztotarget)!($(\tikztostart)!#1!\pgfkeysvalueof{/tikz/ncbar angle}:(\tikztotarget)$)!\pgfkeysvalueof{/tikz/ncbar angle}:(\tikztostart)$)
        -- (\tikztotarget)
    },
    ncbar/.default=0.5cm,
}
\tikzset{round left paren/.style={ncbar=0.5cm,out=110,in=-110}}
\tikzset{round right paren/.style={ncbar=0.5cm,out=70,in=-70}}

\usepackage{hyperref}
\usepackage{url}

\usepackage{amsmath}
\usepackage{amsmath,amsthm}
\usepackage{lmodern}
\usepackage{pgfplots}
\usepackage{algorithm}
\usepackage[noend]{algpseudocode}
\usepackage[export]{adjustbox}

\usepackage{comment}

\begin{document}

\title{Using Homomorphic hashes in coded blockchains}

\author{
\IEEEauthorblockN{Doriane Perard, Xavier Goffin, J\'er\^ome Lacan}
\IEEEauthorblockA{ISAE-Supaero, Universit\'{e} de Toulouse, France\\
firstname.name@isae-supaero.fr}
}

\maketitle

\begin{abstract}
One of the scalability issues of blockchains is the increase of their sizes which can prevent users from storing them and thus from contributing to the decentralization effort. Recent works developed the concept of coded blockchains, which allow users to store only some coded fragments of the blockchains. However, this solution is not protected against malicious nodes that can propagate erroneous coded fragments. 

We propose in the paper to add homomorphic hashes to this system. This allows for instantaneous detection of erroneous fragments and thus avoids decoding with wrong data. We describe the integration of this mechanism in coded blockchains and we evaluate its complexity theoretically and by simulation.  
\end{abstract}

\section{Introduction}
\label{sec:intro}

One of the most interesting properties of blockchains is their decentralized nature, making it possible not to use central authorities. 
Usually, each node participating in the blockchain must maintain it by participating in the consensus when inserting a new block and by storing the entire blockchain. However, the success of blockchains such as Bitcoin or Ethereum has highlighted a scalability problem. Indeed, the increasing size of these blockchains means further constraints for medium-capacity nodes, leading to consequences on the availability and decentralization of the blockchain.

In order to allow nodes with a limited storage capacity to participate in the storage of the blockchain, several works introduced and studied  \emph{coded blockchains} by using erasure codes or network coding. The main principle is to store only some \emph{coded fragments} of each block. These fragments are obtained by first splitting a block into $k$ fixed size fragments and then generating linear combinations of these fragments. These linear combinations can be randomly generated \cite{perard2018erasure} or can follow a structured code such as Low-Density Parity Check (LDPC) codes \cite{wu2020distributed} or Fountain codes \cite{kadhe2019sef}. The average number of source blocks included in a linear combination is called the degree $d$.   

When a node wants to join the network, it needs to download and verify each block of the blockchain, generate coded fragments, and then delete the original blocks to keep its coded fragments only.
When the node wants to restore a block, it downloads $k(1+ \epsilon)$ coded fragments and then performs the reverse operation. The value of the real number $\epsilon$ can vary from $0$ (ideal code) up to $0.5$ according to $d$ and the used code.  

However, these propositions do not consider adversarial nodes that can provide maliciously formed coded fragments and thus prevent the correct decoding of the whole block.

To detect bad coded fragments, we propose in this paper to use \textit{homomorphic hashing} functions that were introduced by \cite{on_the_fly_2004} and improved by \cite{infocomsecurity_2006} in the context of peer-to-peer distributed storage systems. 
The main property of these hashes is that the hash of a linear combination of source blocks can be expressed as a function of the linear combination coefficients and hashes of source blocks. Thus, if the hashes of the source blocks are public and certified, any user can verify the validity of a received block by checking that its hash corresponds to the output of the verification function.


This paper is structured as follows. Sections~\ref{sec:description} presents a description of erasure-code based low storage nodes using homomorphic hashes, the main contribution of this paper. Afterwards, Sections~\ref{sec:interest} and~\ref{sec:analysisParameters} present the interest of our low storage blockchain node and an analysis of the available parameters of our system. Finally, Section~\ref{sec:conclusion} concludes this paper and exposes ways to further this topic.

\section{Including Homomorphic Hashes in Coded Blockchains}
\label{sec:description}

In this article we propose to use homomorphic hashing functions on coded blockchains in order to detect erroneous coded fragments. It will then be possible to replace them by correct ones, and to list malicious nodes.

Before describing the coding operation, let us first define some notations. We denote by $N^{(i)}$ the nodes of the network, where $i$ is an unique identifier characterizing each node. We denote by $B^{(j)}$ the $j^{th}$ block of the blockchain. We consider that the first block is $B^{(0)}$. 

We consider that hashing and coding operations are done on the finite field $\mathbb{Z}_p$, where $p$ is a prime number of size $|p|$ bytes. Let $s_B$ the maximum size of a block of the blockchain (in bytes). Let us define two integers $k$ and $r$ respectively corresponding to the number of fragments of a block and the number of coded fragments stored by a node. The size of a fragment is denoted by $m=s_B/|p|$. The choice of the values of system parameters as $k$, $r$ and $p$ 
will be discussed in Section~\ref{sec:analysisParameters}. 

\subsection{Coding the data}
\label{subsec:coding}


\subsubsection{Block Splitting} The block $B^{(j)}$ is split into $k$ fragments  $F^{(j)}_{l}$, with $l=1,\ldots,k$ . 
The fragments are themselves composed of $m$ finite field elements. The elements of the fragment $F^{(j)}_{l}$ are denoted $f^{(j)}_{l,v}$, where $v=1,\ldots,m$. 

The last fragment can be padded if needed, in case of variations in block size. 

\subsubsection{Homomorphic Hash of the Block} As defined in \cite{on_the_fly_2004}, we consider that the following public parameters define the system 
$G = (p, q, g)$ 
where  $q$ is a large prime number 
such that $q | (p - 1)$. The vector $g$ is composed of some elements  $g_i \in \mathbb{Z}_p$, for $i=1,\ldots,m$. 
The hash of the block $B^{(j)}$ corresponds to the set of the hashes of its fragments : $ h(B^{(j)}) = ( h(F^{(j)}_{1}) , h(F^{(j)}_{2}) , ... , h(F^{(j)}_{k}))$ where: 
$$ h(F^{(j)}_{l})= \prod_{v=1}^{m} g_{v}^{f^{(j)}_{l,v}} \mod \ p $$

\subsubsection{Coded Fragments Generation} 
%
To build the coded fragment $\mathfrak{F}^{(i,j)}_{u}$, where $0\leq~u~\leq~r-1$, the node considers the  $k$ coefficients $\{\alpha^{(i,j)}_{k.u}, \ldots, \alpha^{(i,j)}_{(k+1).u-1} \}$ and computes the following linear combination:
$$ \mathfrak{F}^{(i,j)}_{u} = \alpha^{(i,j)}_{k.u}.F^{(j)}_{1}  
+ \ldots, \alpha^{(i,j)}_{(k+1).u-1} .F^{(j)}_{k}$$

We assume that the values of $\alpha^{(i,j)}_{k.u}+v$ can be deduced from $i$ and $j$. 
From a more practical point of view, 
the $v^{th}$ element of the coded fragment $\mathfrak{F}^{(i,j)}_{u}$ is defined by the finite field element $\mathfrak{f}^{(i,j)}_{u,v}$ computed as follows: 
$$ \mathfrak{f}^{(i,j)}_{u,v} = \alpha^{(i,j)}_{k.u}.f^{(j)}_{1,v}  
+ \ldots, + \alpha^{(i,j)}_{(k+1).u-1}.f^{(j)}_{k,v}$$ 

The  $k$ coefficients $\{\alpha^{(i,j)}_{k.u}, \ldots, \alpha^{(i,j)}_{(k+1).u-1} \}$ depend on the chosen erasure code. 

\begin{figure}[htb]
	\centering
	\scalebox{1}{\resizebox{0.48\textwidth}{!}{
\begin{tikzpicture}[empty/.style={draw,thick,rounded corners,inner sep=.0cm}, to/.style={->,>=stealth',semithick,font=\footnotesize}, dot/.style={dotted,semithick,font=\footnotesize}]

\draw(7.9,-4.5) node[empty,minimum width=90, minimum height=80,rectangle split, rectangle split parts=5, rectangle split draw splits=false, rotate=0] (n1) {
\nodepart{one}
	\parbox[c][1cm]{40pt}{
		\begin{center}
			$F^{(j)}_{0}$
		\end{center}
	} 
\nodepart{two} 
	\parbox[c][1cm]{40pt}{
		\begin{center}
			$F^{(j)}_{1}$
		\end{center}
	} 
\nodepart{three} 
	\parbox[c][1cm]{40pt}{
		\begin{center}
			$F^{(j)}_{2}$ 
		\end{center}
	}
\nodepart{four}	
	\parbox[c][1cm]{40pt}{
		\begin{center} 
			$\ldots$
		\end{center}
	}
\nodepart{five}	
	\parbox[c][1cm]{40pt}{
		\begin{center} 

			$F^{(j)}_{k-1}$
		\end{center}
	}
};

 \draw[dashed] (n1.one split east) -- (n1.one split west);
  \draw[dashed] (n1.two split east) -- (n1.two split west);
   \draw[dashed] (n1.three split east) -- (n1.three split west);
      \draw[dashed] (n1.four split east) -- (n1.four split west);

\draw(0,-3) node[minimum width=35, minimum height=35] (m11) {$\alpha^{(i,j)}_{0}$};
\draw(0,-4) node[minimum width=35, minimum height=35] (m11b) {$\alpha^{(i,j)}_{k}$};
\draw(0,-5) node[minimum width=25, minimum height=35] (m12) {$\ldots$};
\draw(0,-6) node[minimum width=35, minimum height=35] (m13) {$\ldots$};

\draw(0,-3) node[minimum width=35, minimum height=35] (m11) {$\alpha^{(i,j)}_{0}$};
\draw(0,-4) node[minimum width=35, minimum height=35] (m11b) {$\alpha^{(i,j)}_{k}$};
\draw(0,-5) node[minimum width=25, minimum height=35] (m12) {$\ldots$};
\draw(0,-6) node[minimum width=35, minimum height=35] (m13) {$\ldots$};

\draw(1,-3) node[minimum width=25, minimum height=35] (m21) {$\alpha^{(i,j)}_{1}$};
\draw(1,-4) node[minimum width=25, minimum height=35] (m21b) {$\alpha^{(i,j)}_{k+1}$};
\draw(1,-5) node[minimum width=25, minimum height=35] (m22) {$\ldots$};
\draw(1,-6) node[minimum width=25, minimum height=35] (m23) {$\ldots$};

\draw(2,-3) node[minimum width=25, minimum height=35] (m31) {$\alpha^{(i,j)}_{2}$};
\draw(2,-4) node[minimum width=25, minimum height=35] (m31) {$\alpha^{(i,j)}_{k+2}$};
\draw(2,-5) node[minimum width=25, minimum height=35] (m32) {$\ldots$};
\draw(2,-6) node[minimum width=25, minimum height=35] (m33) {$\ldots$};

\draw(3,-3) node[minimum width=25, minimum height=35] (m31) {$\ldots$};
\draw(3,-4) node[minimum width=25, minimum height=35] (m31) {$\ldots$};
\draw(3,-5) node[minimum width=25, minimum height=35] (m32) {$\ldots$};
\draw(3,-6) node[minimum width=25, minimum height=35] (m33) {$\ldots$};

\draw(4,-3) node[minimum width=35, minimum height=35] (m41) {$\alpha^{(i,j)}_{k-1}$};
\draw(4,-4) node[minimum width=35, minimum height=35] (m41b) {$\alpha^{(i,j)}_{2k-1}$};
\draw(4,-5) node[minimum width=25, minimum height=35] (m42) {$\ldots$};
\draw(4,-6) node[minimum width=35, minimum height=35,align=center] (m43) {$\alpha^{(i,j)}_{r.k-1}$};
\draw[thick]
 (m43.south east)to[round right paren] node[yshift=50pt]{} (m41.north east) ;

 \draw[thick]
  (m13.south west)to[round left paren] (m11.north west) ;

\draw(10.5,-4.5) node[] {\LARGE =};

\draw(5.7,-4.5) node[] {\LARGE $\times$};

\draw(13,-4.5) node[empty,minimum width=90, minimum height=80,rectangle split,  rectangle split parts=4, rectangle split draw splits=false, rotate=0] (n1) {
\nodepart{one}
	\parbox[c][1cm]{40pt}{
		\begin{center}
			$\mathfrak{F}^{(i,j)}_{0}$
		\end{center}
	} 
\nodepart{two} 
	\parbox[c][1cm]{40pt}{
		\begin{center}
			$\mathfrak{F}^{(i,j)}_{1}$
		\end{center}
	} 
\nodepart{three} 
	\parbox[c][1cm]{40pt}{
		\begin{center}
			$\ldots$
		\end{center}
	} 
\nodepart{four} 
	\parbox[c][1cm]{40pt}{
		\begin{center}
			$\mathfrak{F}^{(i,j)}_{r-1}$
		\end{center}
	}
};
 \draw[dashed] (n1.one split east) -- (n1.one split west);
  \draw[dashed] (n1.two split east) -- (n1.two split west);
    \draw[dashed] (n1.three split east) -- (n1.three split west);

  \draw(13,-1.5) node[fill=black, rectangle callout, callout absolute pointer={(12,-3)}] {Coded fragment};
  \draw(13,-1.5) node[empty,fill=gray!20, rounded rectangle, text width=3cm, minimum height=20pt, align=center] {Coded fragment};

  \draw(9.5,-1.0) node[fill=black, rectangle callout, callout absolute pointer={(8.8,-2.5)}] {Fragment};
   \draw(9.5,-1.0) node[empty,fill=gray!20, rounded rectangle, text width=2.5cm, minimum height=20pt, align=center] {Fragment};

  \draw(5.4,-1.0) node[fill=black, rectangle callout, callout absolute pointer={(7.0,-1.95)}] {Block $B^{(j)}$};
    \draw(5.4,-1.0) node[empty,fill=gray!20, rounded rectangle, text width=2.5cm, minimum height=20pt, align=center] {Block $B^{(j)}$};

\end{tikzpicture}}}
	\caption{Block coding process}
	\label{fig:bcCoded}
\end{figure}
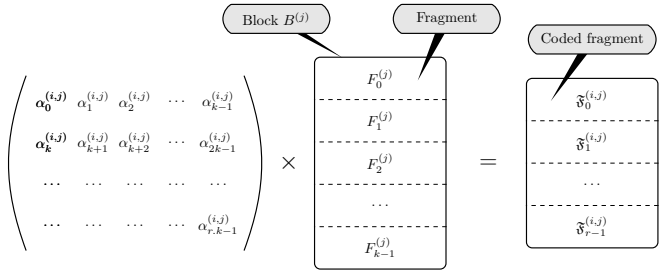

\subsubsection{Hash of the Coded Fragments} The hash of each of the $r$ coded fragments is then computed. 
We denote it by $\mathfrak{h}_u^{(i,j)} = h( \mathfrak{F}^{(i,j)}_u )$. Thanks to the homomorphic property, it can be proved that 
\begin{equation}
 h( \mathfrak{F}^{(i,j)}_u ) = \prod_{v=1}^{k} h(  F_v^{(j)} )^{\alpha^{(i,j)}_{k.u+v-1}} 
 \label{eq:homomorphic}
\end{equation}

\subsubsection{Storing Data} Finally, the node removes the block $B^{(j)}$ and replaces it by the $r$ coded fragments $\mathfrak{F}^{(i,j)}_{1},\ldots,\mathfrak{F}^{(i,j)}_{r}$, their $r$ hashes and the $k$ hashes of the initial fragments.

\subsection{Recovering the data}
\label{subsec:recovery}
\begin{figure}[htb]
	\centering
	\scalebox{.8}{\tikzset{every picture/.style={line width=0.75pt}} 

\begin{tikzpicture}[x=0.75pt,y=0.75pt,yscale=-1,xscale=1]

\draw   (30,47) -- (140.5,47) -- (140.5,77) -- (30,77) -- cycle ;
\draw   (161,52.9) .. controls (161,49.64) and (163.64,47) .. (166.9,47) -- (195.1,47) .. controls (198.36,47) and (201,49.64) .. (201,52.9) -- (201,70.6) .. controls (201,73.86) and (198.36,76.5) .. (195.1,76.5) -- (166.9,76.5) .. controls (163.64,76.5) and (161,73.86) .. (161,70.6) -- cycle ;
\draw   (31,157) -- (139.5,157) -- (139.5,188) -- (31,188) -- cycle ;
\draw   (160,162.3) .. controls (160,158.82) and (162.82,156) .. (166.3,156) -- (193.7,156) .. controls (197.18,156) and (200,158.82) .. (200,162.3) -- (200,181.2) .. controls (200,184.68) and (197.18,187.5) .. (193.7,187.5) -- (166.3,187.5) .. controls (162.82,187.5) and (160,184.68) .. (160,181.2) -- cycle ;
\draw   (30,207) -- (139.5,207) -- (139.5,237) -- (30,237) -- cycle ;
\draw   (160,213.4) .. controls (160,210.14) and (162.64,207.5) .. (165.9,207.5) -- (193.1,207.5) .. controls (196.36,207.5) and (199,210.14) .. (199,213.4) -- (199,231.1) .. controls (199,234.36) and (196.36,237) .. (193.1,237) -- (165.9,237) .. controls (162.64,237) and (160,234.36) .. (160,231.1) -- cycle ;
\draw   (253,47) -- (361,47) -- (361,76) -- (253,76) -- cycle ;
\draw   (372,52.9) .. controls (372,49.64) and (374.64,47) .. (377.9,47) -- (405.1,47) .. controls (408.36,47) and (411,49.64) .. (411,52.9) -- (411,70.6) .. controls (411,73.86) and (408.36,76.5) .. (405.1,76.5) -- (377.9,76.5) .. controls (374.64,76.5) and (372,73.86) .. (372,70.6) -- cycle ;
\draw   (250,156) -- (361,156) -- (361,188) -- (250,188) -- cycle ;
\draw   (372,162.3) .. controls (372,158.82) and (374.82,156) .. (378.3,156) -- (404.7,156) .. controls (408.18,156) and (411,158.82) .. (411,162.3) -- (411,181.2) .. controls (411,184.68) and (408.18,187.5) .. (404.7,187.5) -- (378.3,187.5) .. controls (374.82,187.5) and (372,184.68) .. (372,181.2) -- cycle ;
\draw    (195.1,76.5) .. controls (210.8,87) and (189,185) .. (243,193) .. controls (296.19,200.88) and (351.81,219.43) .. (377.35,196.1) ;
\draw [shift={(378.5,195)}, rotate = 495] [color={rgb, 255:red, 0; green, 0; blue, 0 }  ][line width=0.75]    (10.93,-3.29) .. controls (6.95,-1.4) and (3.31,-0.3) .. (0,0) .. controls (3.31,0.3) and (6.95,1.4) .. (10.93,3.29)   ;
\draw    (193.7,187.5) .. controls (209.4,198) and (190.5,197) .. (244.5,205) .. controls (297.69,212.88) and (349.91,235.31) .. (375.36,212.09) ;
\draw [shift={(376.5,211)}, rotate = 495] [color={rgb, 255:red, 0; green, 0; blue, 0 }  ][line width=0.75]    (10.93,-3.29) .. controls (6.95,-1.4) and (3.31,-0.3) .. (0,0) .. controls (3.31,0.3) and (6.95,1.4) .. (10.93,3.29)   ;
\draw    (199,231.1) .. controls (218,234) and (193.5,229) .. (247.5,237) .. controls (300.69,244.88) and (349.03,251.79) .. (374.36,228.1) ;
\draw [shift={(375.5,227)}, rotate = 495] [color={rgb, 255:red, 0; green, 0; blue, 0 }  ][line width=0.75]    (10.93,-3.29) .. controls (6.95,-1.4) and (3.31,-0.3) .. (0,0) .. controls (3.31,0.3) and (6.95,1.4) .. (10.93,3.29)   ;
\draw    (405.1,76.5) .. controls (439.4,79) and (466,256) .. (421,257) .. controls (376.67,257.99) and (350.78,264.79) .. (373.91,241.1) ;
\draw [shift={(375,240)}, rotate = 495] [color={rgb, 255:red, 0; green, 0; blue, 0 }  ][line width=0.75]    (10.93,-3.29) .. controls (6.95,-1.4) and (3.31,-0.3) .. (0,0) .. controls (3.31,0.3) and (6.95,1.4) .. (10.93,3.29)   ;

\draw (68,50.9) node [anchor=north west][inner sep=0.75pt]    {$\mathfrak{F}^{( 1,j)}_{1}$};
\draw (69,159.9) node [anchor=north west][inner sep=0.75pt]    {$\mathfrak{F}^{( 3,j)}_{1}$};
\draw (70,209.9) node [anchor=north west][inner sep=0.75pt]    {$\mathfrak{F}^{( 3,j)}_{2}$};
\draw (293,49.9) node [anchor=north west][inner sep=0.75pt]    {$\mathfrak{F}^{( 2,j)}_{1}$};
\draw (293,159.9) node [anchor=north west][inner sep=0.75pt]    {$\mathfrak{F}^{( 4,j)}_{1}$};
\draw (166,52.4) node [anchor=north west][inner sep=0.75pt]    {$\mathfrak{h}^{( 1,j)}_{1}$};
\draw (164,163.4) node [anchor=north west][inner sep=0.75pt]    {$\mathfrak{h}^{( 3,j)}_{1}$};
\draw (165.9,210.9) node [anchor=north west][inner sep=0.75pt]    {$\mathfrak{h}^{( 3,j)}_{2}$};
\draw (375,52.4) node [anchor=north west][inner sep=0.75pt]    {$\mathfrak{h}^{( 2,j)}_{1}$};
\draw (375,163.4) node [anchor=north west][inner sep=0.75pt]    {$\mathfrak{h}^{( 4,j)}_{1}$};
\draw (132,9.4) node [anchor=north west][inner sep=0.75pt]  [font=\Large]  {$N^{( 1)}$};
\draw (129,123.4) node [anchor=north west][inner sep=0.75pt]  [font=\Large]  {$N^{( 3)}$};
\draw (346,12.4) node [anchor=north west][inner sep=0.75pt]  [font=\Large]  {$N^{( 2)}$};
\draw (352,119.4) node [anchor=north west][inner sep=0.75pt]  [font=\Large]  {$N^{( 4)}$};

\end{tikzpicture}}
	\caption{Homomorphic hashes verification }
	\label{fig:systeme_hash_homomorphe}
\end{figure}
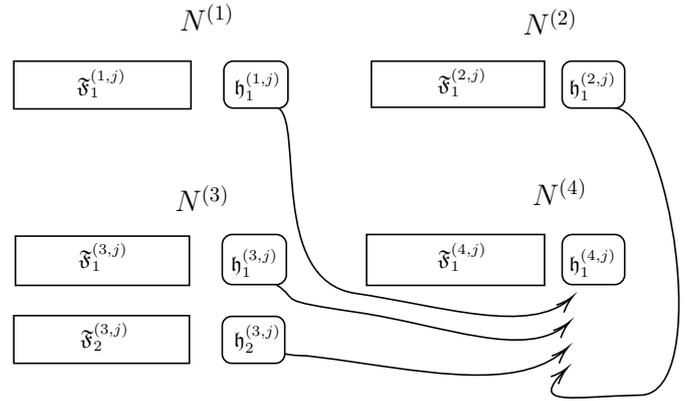
%

When a Low Storage node (LS node) wants to recover a block from coded fragments stored by different LS nodes, it executes the following steps illustrated on Fig.~\ref{fig:systeme_hash_homomorphe}.



\subsubsection{Download the Coded Fragments Hashes} 
As represented in Fig.~\ref{fig:systeme_hash_homomorphe}, node $N^{(4)}$ first downloads $k(1+\epsilon)$ hashes $\mathfrak{h}_z^{(i,j)} $ from different nodes. 

\subsubsection{Hash Check} The node checks that the downloaded hashes are correct by verifying Eq. \ref{eq:homomorphic}. Recall that the values of $\alpha^{(i,j)}_{k.u}$ can be deduced from $i$ and $j$.


\subsubsection{Download the Coded Fragments} If the hashes are verified, the node $N^{(i)}$ downloads coded fragments associated with the previous hashes $\mathfrak{F}^{(i_{0},j)}_{0},\ldots,\mathfrak{F}^{(i_{r-1},j)}_{r-1}$ from several nodes $N^{(i_0)},\dots,N^{(i_{r-1})}$. Note that it is possible to request multiple coded fragments from the same node

\subsubsection{Coded fragment Check} The node hashes each received coded fragment to verify that it matches the corresponding received hash.  

\subsubsection{Block Decoding} Once the fragment hashes are verified, the block can be decoded.
After downloading a sufficient number of coded fragments, the node will have enough equations to invert the linear system and recover the $k$ fragments (and thus the block) from the coded fragments. 


\section{Secure Low storage node interests}
\label{sec:interest}
The main objective of traditional LS nodes is to allow any node to contribute to an entire blockchain with a reduced storage effort. The addition of homomorphic hashes increases the security of the distributed coding process and allows for the identification of malicious nodes. 


\subsection{Scalability}
\subsubsection{Storage effort scalability}
\label{sec:sizes}

With traditional coded blockchain, a node only store $r/k$ data form each block. Let's define the compression factor $c = r/k$.

One of the interests of our system is its scalability. Indeed, each node can adapt $r$ according to, for example, the age of a block by simply removing some of its stored coded fragments without re-calculating them.

Moreover, the number of coded fragments generated and stored on each node can independently be defined by each node, and should be adapted according to the desired storage effort of each node.

\subsubsection{Availability}
One of the main goals of our system is to improve the global availability  and  sustainability  of  a  blockchain. By reducing the storage effort needed to participate, we expect more participants storing at least one coded fragment of every block.  
This means that for a system with a large amount of nodes, any node can then leave the system or  be  unreachable  without  significantly  impacting  the availability.




\subsubsection{Network improvement}
With the increase of the amount of nodes, we improve the distribution of the blockchain over the network. Our low storage nodes can allow for the decongestion of the network. 

\subsection{Malicious  node  identification}
With homomorphic hashing, it becomes possible to identify malicious nodes in the network, providing incorrect coded fragments. A simple solution to avoid them is to locally blacklist them and avoid contacting them in the future. 

But we can also imagine a network level impact, where cheaters are publicly denounced. An incentive system can be easily set up, by punishing malicious nodes and rewarding senders of valid denunciations.    
It can be done by using fraud proofs system, as presented in~\cite{al2018fraud}.

\section{Analysis of the parameters}
\label{sec:analysisParameters}

One of the challenges is to determine $k$, $r$ and the security parameters, with the best compromise between compression, complexity and security. In this section, we will present some consequences when varying these parameters. 


\subsection{Type and size of the finite field}
\label{sec:sizeFiniteField}
The linear combinations of the code and the hash operations are performed on finite fields. Practically, the data of the blocks are grouped into bit vectors of fixed length which are associated to finite field elements and processed with the corresponding rules. The homomorphic property of the hash implies that the code and the hash use the same finite field. With the considered type of hash, a finite field of type $\mathbb{Z}_p$, where the operations are performed modulo a large prime number $p$ must be used. 

The choice of the finite field impacts the probability of block recovery from downloaded coded fragments (which is better with a large finite field) and the encoding and decoding complexities (which is smaller with a small finite field). The size of the finite field is also a security parameter because a minimal value is necessary to avoid collisions. Under these constraints, the choice of a value of $p$ with $1024-bit$ length is chosen, as suggested in \cite{on_the_fly_2004}.

\subsection{Processing coding complexity}
\label{sec:procComplexity}
The complexity of encoding consists in multiplying a $k \times r$-matrix by the $k$ original fragments. Then, there is $d \times r \times s_B/k$ operations in the finite field, so when $d=k$, the encoding complexity is $r \times s_B $. So, the encoding complexity does not depend on $k$, but only on $r$. 

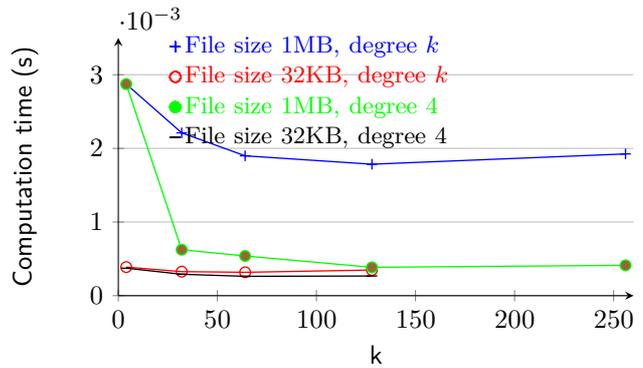
\begin{figure}
\begin{tikzpicture}
\centering
\begin{axis}[
	font=\sffamily,
	height=5cm,
	width=\linewidth-0.5cm,
	axis x line=bottom,
	axis y line=left,
	ymajorgrids,
	xlabel = {k} ,
	ylabel = {Computation time (s)},
	xmin = 0, xmax =260, 
	ymin = 0, ymax = 0.0035 ]
\addplot+[blue, mark=+, line width=0.5pt] table[x index=1, y index=0] {data/data_aux_block_time_1Mo_degrek.txt};
\addplot+[red, mark=o, line width=0.5pt] table[x index=1, y index=0] {data/data_aux_block_time_32ko_degrek.txt};
\addplot+[green, mark=*, line width=0.5pt] table[x index=1, y index=0] {data/data_aux_block_time_1Mo_degre4.txt};
\addplot+[black, mark=-, line width=0.5pt] table[x index=1, y index=0] {data/data_aux_block_time_32ko_degre4.txt};
\end{axis}
\begin{scope}[shift={  (0.5,3.5)  }]
	\draw[xshift=0cm, yshift=-0.2cm, blue] (0,0) 
	plot[mark=+] (0.25,0)
	node[right]{\small{File size 1MB, degree $k$}};
	\draw[xshift=0cm, yshift=-0.6cm, red] (0,0)
	plot[mark=o] (0.25,0) 
	node[right]{\small{File size 32KB, degree $k$}};
    \draw[xshift=0cm, yshift=-1.0cm, green] (0,0) 
	plot[mark=*] (0.25,0)
	node[right]{\small{File size 1MB, degree $4$}};
	\draw[xshift=0cm, yshift=-1.4cm, black] (0,0)
	plot[mark=-] (0.25,0) 
	node[right]{\small{File size 32KB, degree $4$}};

\end{scope}	

\end{tikzpicture}
\caption{Coded fragment time generation, from different size file and $k$.}
\label{fig:coding_speed}
\end{figure}

Fig.~\ref{fig:coding_speed} shows the encoding speeds of 1MB and 32 kB blocks, with different values of degres and $k$. The implemented code was run in a Virtual Machine with operating system Debian 10. This personal computer runs Windows 10 and is equipped with an Intel Core i5-7300HQ @2.50GHz with  8GB of RAM.
This graph allows to conclude that the processing cost is acceptable. Indeed, coding speed is always around milliseconds.

For decoding, the complexity consists in inverting a $k \times k$-matrix, and then multiplying it by the $k$ coded fragments. The pseudo-random matrix inversion has a complexity in $O(k^3)$. So the number of operations is ${O(k^3) + k^2 \times s_B/k} {= O(k^3) + k \times s_B} $ and thus depends only on $k$.
If the size of the block is large compared to $k$, then the matrix-vector multiplication ($k\times$ block size) is the most complex operation.

\subsection{Homomorphic hashing functions complexity and parameters}

\begin{figure}
\begin{tikzpicture}
\centering

    \begin{semilogyaxis}[
	font=\sffamily,
	height=5cm,
	width=\linewidth-0.5cm,
	axis x line=bottom,
	axis y line=left,
	ymajorgrids,
	xlabel = {k} ,
	ylabel = {Computation time (s)},
	xmin = 0, xmax =260, 
	ymin = 0, ymax = 0.25 ]
\addplot+[blue, mark=+, line width=0.5pt] table[x index=2, y index=0] {data/data_aux_block_gene_verif_1Mo_degrek.txt};
\addplot+[red, mark=o, line width=0.5pt] table[x index=2, y index=0] {data/data_aux_block_gene_verif_32ko_degrek.txt};
    \end{semilogyaxis}
    
\begin{scope}[shift={  (0.5,3.5)  }]
	\draw[xshift=0cm, yshift=-0.2cm, blue] (0,0) 
	plot[mark=+] (0.25,0)
	node[right]{\small{File size 1MB}};
	\draw[xshift=0cm, yshift=-0.6cm, red] (0,0)
	plot[mark=o] (0.25,0) 
	node[right]{\small{File size 32KB}};

\end{scope}	
\end{tikzpicture}
\caption{Time to hash a fragment}
\label{fig:verif2}
\end{figure}
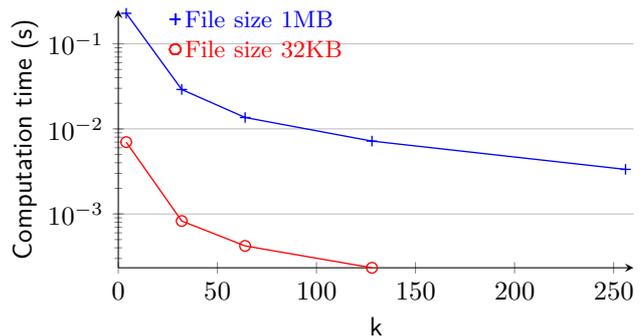

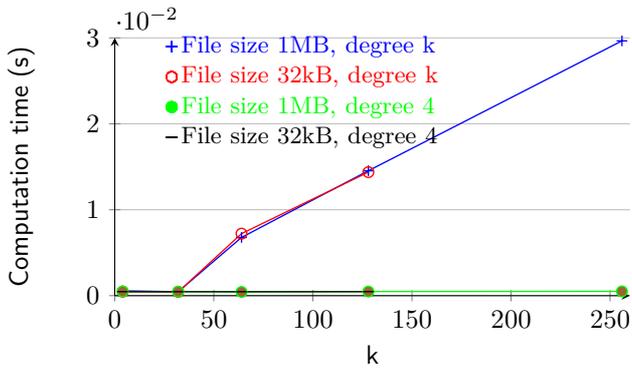
\begin{figure}
\begin{tikzpicture}
\centering
\begin{axis}[
	font=\sffamily,
	height=5cm,
	width=\linewidth-0.5cm,
	axis x line=bottom,
	axis y line=left,
	ymajorgrids,
	xlabel = {k} ,
	ylabel = {Computation time (s)},
	xmin = 0, xmax =260, 
	ymin = 0, ymax = 0.03 ]
\addplot+[blue, mark=+, line width=0.5pt] table[x index=2, y index=1] {data/data_aux_block_gene_verif_1Mo_degrek.txt};
\addplot+[red, mark=o, line width=0.5pt] table[x index=2, y index=1] {data/data_aux_block_gene_verif_32ko_degrek.txt};
\addplot+[green, mark=*, line width=0.5pt] table[x index=2, y index=1] {data/data_aux_block_gene_verif_1Mo_degre4.txt};
\addplot+[black, mark=-, line width=0.5pt] table[x index=2, y index=1] {data/data_aux_block_gene_verif_32ko_degre4.txt};
\end{axis}
\begin{scope}[shift={  (0.5,3.5)  }]
	\draw[xshift=0cm, yshift=-0.2cm, blue] (0,0) 
	plot[mark=+] (0.25,0)
	node[right]{\small{File size 1MB, degree k}};
	\draw[xshift=0cm, yshift=-0.6cm, red] (0,0)
	plot[mark=o] (0.25,0) 
	node[right]{\small{File size 32kB, degree k}};
    \draw[xshift=0cm, yshift=-1.0cm, green] (0,0) 
	plot[mark=*] (0.25,0)
	node[right]{\small{File size 1MB, degree 4}};
	\draw[xshift=0cm, yshift=-1.4cm, black] (0,0)
	plot[mark=-] (0.25,0) 
	node[right]{\small{File size 32kB, degree 4}};

\end{scope}	

\end{tikzpicture}
\caption{Time to generate an homomorphic hash of a coded fragment from original fragment hashes}
\label{fig:verif1}
\end{figure}

According to \cite{on_the_fly_2004} and \cite{infocomsecurity_2006}, the complexity to hash a $m$-element fragment is  $O(m)$, and thus a file of $k$ fragments is hashed in  $O(k.m)$. Fig.~\ref{fig:verif2} confirms that, because when $k$ increases, the fragment size $m$ decreases, and so does the time.

To check the validity of the hash of a coded fragment from $d$ source hashes, the complexity is $O(m) + O(d)$. Fig.~\ref{fig:verif1} shows us that the fragment size $m$ is not so important comparing to the degree $d$, in terms of complexity. When this degree is low, for example 4, the time is low too (around 0.0004 s). But when it is equal to $k$ (i.e. coded fragments are composed by linear combinations of every fragments), the time increases when $k$ does. 

The parameters are defined at the system level and are therefore the same on all nodes. This choice is important, because it will have a direct influence on the level of security but also on the complexity of the operations to be performed. 

The time to perform homomorphic hash is independent of $k$. 

\subsection{Compression factor}
\label{subsection:compression_factor}

With the homomorphic hashes system, the compression factor $c$ changes. In this system we have to store extra data : homomorphic hashes of all the original fragments and homomorphic hashes of all the coded fragments. The new formula is so:

\begin{equation}
    c = \frac{(k+r) \times S_H}{S_B} + \frac{r}{k}
    \label{eq:compression_factor}
\end{equation}

To find the optimum of this equation, we can calculate:  
\begin{equation}
    k_{opt} = \sqrt{\frac{r \times S_B}{S_H}} 
    \label{eq:find_the_better_k}
\end{equation}

\begin{table}[]
    \centering
    \begin{tabular}{|c|c|c|c|c|c|}
    \hline
         \textbf{k} & \textbf{4} & \textbf{32} & \textbf{64} & \textbf{128} & \textbf{256}  \\
    \hline
         \textbf{r = 1} & 0.251 & 0.0355 & 0.0239 & 0.0243 & 0.0369 \\
    \hline
        \textbf{ r = 5} & - & 0.161 & 0.0870 & 0.0561 & 0.0529 \\
    \hline
    \end{tabular}
    \caption{Values of $c$ for variation of $k$ and $r$, with $s_B = 1 MB$}
    \label{tab:compression_factors}
\end{table}

\subsection{Chosing $k$ and $r$}

As described in Section~\ref{subsec:coding}, each node, in order to generate its $r$ coded fragments, will split the initial block into $k$ fragments. The choice of this parameter can be different for each blockchain, but it must be the same for every user of the same blockchain. 

When $k$ increases, there is no impact on block hashing time and encoding speed, but in the end the nodes need to verify more coded fragments, so it will be longer.

The choice of $r$ is up to the end user and will depend on the type of user. Choosing a large $r$ will improve block recovery and reduce network load, as well as improve the overall blockchain availability as increasing $r$ increases the storage effort of a node. It also improves the recovery block speed, because the nodes will verify less $(k+\epsilon-r)$ hashes and coded fragments.
Choosing a small $r$ will reduce coding complexity and compression factor. Globally $r$ must be chosen according to the node's capacities. 

If we want a better compression factor, we can use the formula \ref{eq:find_the_better_k}, with $r=1$, and we can calculate the optimal $k$. If we want an even better compression factor, we can also increase $s_B$ by grouping some blocks before coding them. But during decoding, we will reconstruct more data than we need. 

The Table.~\ref{tab:compression_factors} presents some compression factors for different $k$ and $r$ values, for block size equal to 1MB like in Bitcoin. According to Eq.~\ref{eq:find_the_better_k}, the higher factor compression is 0.228, when $k = 88$ for $r=1$, and 0.0512, when $k=198$ for $r=5$.

\section{Conclusion}
\label{sec:conclusion}
The main contribution of this paper is to introduce homomorphic hashes in coded blockchains. We explained how to compute, store and exchange these hashes in order to detect erroneous coded fragments. The impact of this mechanism in terms of additional storage and complexity was analyzed. A global analysis of the parameters was proposed in order to determine the parameters of the system. 
Future work will focus on the optimization of the parameters according to the considered blockchains and the types of nodes in order to find the best compromise between compression, complexity and security.

\bibliographystyle{IEEEtran}
\bibliography{main}

\end{document}